# Learning Aided Auctioning based Spectrum Access System in a Wireless Optical Network

Atri Mukhopadhyay, Goutam Das

*Abstract*— This paper focusses on Service Level Agreement (SLA) based end-to-end Quality of Service (QoS) maintenance across a wireless optical integrated network. We use long term evolution (LTE) based spectrum access system (SAS) in the wireless network and the optical network is comprised of an Ethernet Passive Optical Network (EPON). The proposal targets a learning-based intelligent SAS where opportunistic allocation of any available bandwidth is done after meeting the SLA requirements. Such an opportunistic allocation is particularly beneficial for nomadic users with varying QoS requirements. The opportunistic allocation is carried out with the help of Vickrey-Clarke-Groves (VCG) auction. The proposal allows the users of the integrated network to decide the payment they want to make in order to opportunistically avail bandwidth. Learning automata is used for the users to intelligently converge to the optimal payment value based on the network load. The payment made by the users is later used by the optical network units of the EPON to prepare the bids for the auction. The proposal has been verified through extensive simulations.

## I. Introduction

THE last decade has witnessed an exponential increase in the wireless spectrum demands due to the wide-spread usage of hand-held networking devices. Unfortunately, the available wireless spectrum is limited and the proposed mitigation scheme towards shifting to the higher frequency bands (typically 10 GHz and above) results in high signal attenuation [1]. Hence, a dynamic spectrum access (DSA) technique like the three-tier spectrum sharing model known as the spectrum access system (SAS) is suggested to be an effective method to maximize the spectrum utilization [2]. The SAS targets spectrum sharing in the 3.5 GHz citizens broadband radio service (CBRS) band [2][3].

The SAS maintains a geo-location database with well-defined exclusion zones, and manages

A. Mukhopadhyay and G. Das are with G. S. Sanyal School of Telecommunications, Indian Institute of Technology, Kharagpur, India. E-mail: atri.mukherji11@gmail.com, gdas@gssst.iitkgp.ac.in.

spectrum sharing in a way that incumbent operations are guaranteed interference protection according to the terms of their assignments whenever they are present in deployed areas [3]. The incumbents are the high priority CBRS users, and they comprise the highest tier. The secondary users are further classified into two tiers; called second tier priority access license (PAL) users and the lower tier general authorized access (GAA) opportunistic users [1]. The PAL users, which are generally the mobile network operators (MNOs), are protected from the GAA users whereas the GAA users do not get any interference protection guarantees. The GAA users need to be actively managed to provide interference protection to the PAL users [1]. Please note that, we only concentrate on the spectrum sharing between PAL and GAA users by assuming that the incumbents are already protected. The interference mitigation for PAL users must be achieved through sharing of minimum information between the GAA and PAL users as the cellular operators might be reluctant in sharing detailed system information. Further, a large complex system relying on real-time information for interference mitigation may malfunction even with slight delay of reception of a vital information [4].

Finally, PAL user agreements are made in the granularity level of the MNOs. The GAA users, on the other hand, do not have any well-defined agreements and hence, they are essentially nomadic users (international roaming users). In networking, a nomadic user does not require a long-term service level agreement (SLA) like the incumbents or the PAL users.

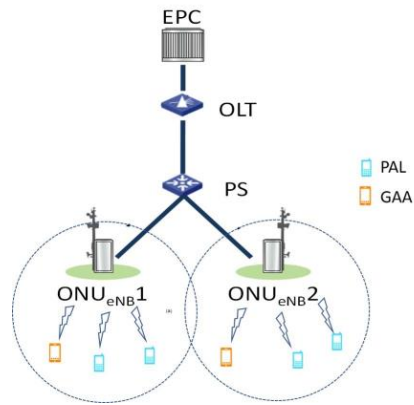

Fig. 1 WOIN architecture

In most of the papers, GAA users are considered as cognitive users where it requires interference management techniques in the physical layer [4]. However, this can be easily avoided if we associate GAA users with the PAL base station (BS) (see Fig. 1) and assume that the MNO per-form scheduling of the GAA users on a competitive pay-as-you-go basis. Here, the PAL BS keeps the provision of scheduling on-demand traffic from GAA users on short term basis. This will also ensure that the GAA users do not interfere with the PAL users.

Therefore, a highly flexible SAS is required to efficiently manage the GAA users. Hence, we move on to propose an advanced version of SAS that is managed from the medium access control (MAC) layer. The proposal governs the network operation using an intelligent opportunistic spectrum sharing scheme. The proposal is software controlled and therefore, can be implemented without the requirement of any specialized hardware. Hence, the overall cost of the network is also reduced. In our proposal, we introduce an artificial intelligence (AI) enabled software defined network (SDN) controller that utilizes learning automata (LA) [7] for resource scheduling among PAL and GAA users to achieve network efficiency goals while satisfying quality of service (QoS) demands [5][6].

After employing SAS as a wireless interface, the network operators need to overcome the issue of routing the huge incoming traffic from the wireless network to the network core. An optical backhaul comes up with the perfect answer to this problem. Thus, we get a wireless optical integrated network (WOIN) illustrated in Fig. 1, where an optical access network is used as the backhaul and a wireless network technology provides the final connectivity [10]-[14]. WOINs utilize the data carrying potential of the optical networks and the ubiquity of the wireless networks. The primary choice for the optical part of the network is Ethernet/Gigabit passive optical network (EPON/GPON) and for the wireless last mile, long term evolution/long term evolution-advanced (LTE/LTE-A)[2] is considered to be the perfect choice [11]. We additionally incorporate an intelligent SAS along with LTE/LTE-A in our proposal.

The EPON/GPON segment of the WOIN is a point-to-multipoint access network with no active elements in its distribution network. An EPON/GPON consists of an optical line terminal (OLT) located at a central office, a passive splitter (PS) and optical network units (ONUs) located in the user premises (Fig. 1). A multipoint control protocol (MPCP) based interleaved polling with adaptive cycle time (IPACT) is used as a statistical multiplexing technique for uplink [9]. The LTE network, on the other hand, consists of base stations (eNodeBs) and user equipments (UEs). Accommodating the varied network traffic requirements of the different network users (both PALs and GAAs) is particularly challenging in WOINs due to its heterogeneity in network

---

[2] We will be referring to both LTE and LTE-A networks as LTE from here onwards.

sub parts.

In order to truly bring out the benefits of the integrated network, the SAS in the mobile network must be complemented by the EPON/GPON backhaul. Normally, bandwidth is over-provisioned while designing the EPON/GPON. Therefore, the higher traffic influx due to our proposed opportunistic wireless spectrum sharing scheme can be easily extended to the EPON/GPON. Thus, we also propose an end-to-end opportunistic bandwidth allocation scheme in this paper. We point out the challenges and solution methodology for the problem in hand in the following sub-sections.

*A. Challenges*

*A.1. Payment Based Fair and Opportunistic Scheduling*

The spectrum sharing mechanism must guarantee the SLA of the PAL users. The remaining bandwidth is allocated to the nomadic GAA users with the help of a fair competitive scheme. We opt for a payment-based competition in our spectrum sharing method.

In a spectrum sharing scenario, it is always desirable that the users (both PAL and GAA) are given more flexibility for deciding the fate of their own traffic. Even in a centralized spectrum sharing mechanism, the users should have the freedom of the payment that they want to offer for the transmission of their traffic.

Normally, an EPON is designed to deal with traffic that vary with time. However, in the present diverse user scenario, a user must be given admission to the network with an SLA. A PAL user has a long-term SLA while the GAA users are not assured of any data rate guarantee by the network. Further, we keep the provision that PAL users might also request for on-demand applications that are not covered by the SLA. Therefore, it must be ensured that if the generated traffic rate from a user exceeds the SLA, then the excess traffic is transmitted only when the user is willing to pay an additional amount. The excess traffic is the additional traffic required to support the on-demand applications. Thus, the user will have the option to decide the value of the additional payment depending on the importance of the packets in the user's buffer. On the other hand, if SLA of any user is violated, the EPON should be imposed with a penalty. Thereby, the WOIN will also be conservative in bandwidth distribution.

Finally, if it is agreed upon that a higher price must be paid for the packets that are above the SLA, the user fairness should also be maintained. Let us consider two scenarios.

1. *Scenario A* – The network load is low and only a single user is operating above SLA.
2. *Scenario B* – The network load is high, and more than one user is operating above SLA and the network can provide extra bandwidth to a subset of users.

In scenario A, the user should be allocated bandwidth in exchange of a minimum cost; since, there is surplus bandwidth available. However, Scenario B is a competitive situation where several users must fight for a limited resource. In such a case, the bandwidth should be allocated to the subset of users paying the highest amounts. Therefore, the network allocates bandwidth to the highest paying users opportunistically depending on the scenario.

Unfortunately, for a user to decide on the fair payment value, information on the overall network load is required. By user fairness, it is intended to convey that if the network load is low and a user is requesting higher bandwidth than mentioned in its SLA, then the payment per byte by the user should be lower than that of the payment requirement for a higher network load. In Fig. 2, we summarize the types of user present in the SAS, the types of traffic that they generate and the category of payments that they make for each traffic types.

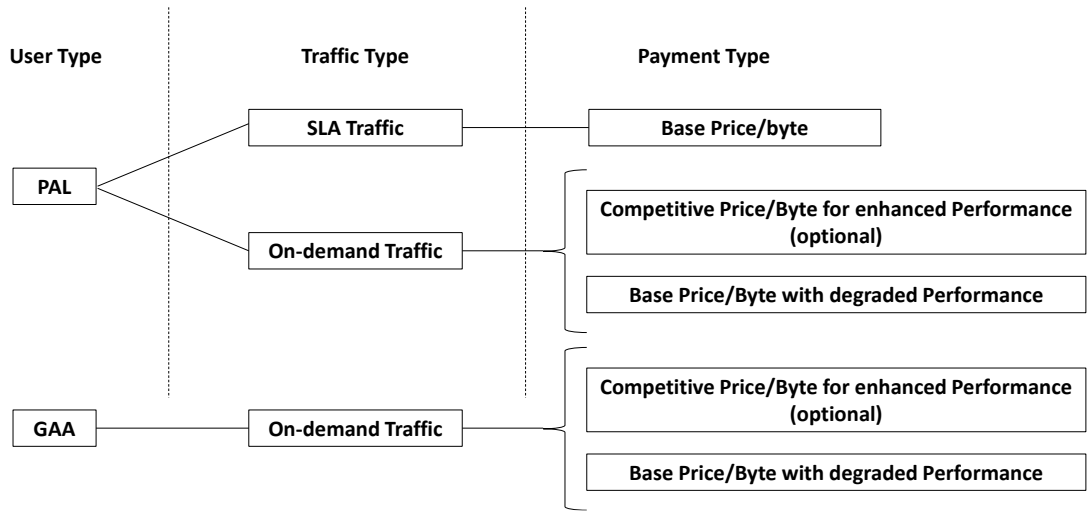

Fig. 2 User Types, their traffic and their payments

*A.2. Incomplete Network Load Information in WOIN Uplink*

In WOIN uplink, the packets from the UEs of the LTE network, reach the OLT of EPON via the ONUs. An eNodeB is directly connected to an ONU in a WOIN to form a composite unit called $ONU_{eNB}$[3]. Therefore, each ONU acts as a traffic aggregation point. In such a scenario, the transmission of a packet from a UE is not only affected by the other UEs directly connected to the same LTE network but also the UEs that are associated to other LTE networks/other ONUs of the same EPON.

---

[3] We will be calling $ONU_{eNB}$ as ONU interchangeably from this point.

Unfortunately, an ONU has no direct means of knowing the overall network load as it cannot interact with the other ONUs. Therefore, any user connected to a certain ONU via its LTE network has no efficient means of perceiving the overall network load. The only way by which the vital traffic information might be exchanged over the WOIN network shall require huge overhead, which is not practical in an already constrained network. Hence, the users must devise some intelligent means to get around the problem and learn the network conditions over time.

We note that multiple entities (UEs) are trying to converge to an optimal decision without interacting with each other. Therefore, it is evident that a distributed learning mechanism is required. On the other hand, the OLT performs uplink scheduling in the EPON without having detailed knowledge about the wireless traffic flows. Hence, in order to design an efficient end-to-end scheduling in the WOIN, we must ensure proper orchestration between the scheduling entities (SDN controllers) like the OLT in the EPON and the eNodeBs in the wireless network.

*B. Solutions*

*B.1. Learning Automata to the Rescue*

Learning Automata is a stochastic learning technique, which learns the optimal action through repeated interactions with its environment. It is therefore a perfect candidate for the given situation. LA is a distributed learning algorithm. Hence, it can be independently executed in each UE. However, implementing LA in a UE is not recommended as it demands high processing power. Therefore, multi-access edge computing (MEC) may be employed where a MEC server hosted inside an eNodeB runs the algorithm on behalf of the UEs with the help of the information received from the UEs [15].

Using LA, the UE learns the network conditions over time. The users can treat the actions of other users as environmental effect and thus observe the outcome against every action that it takes. Slowly, the UE will converge towards an equilibrium decision. This procedure has a finite convergence time. Luckily, in any practical time varying network, a load scenario remains approximately constant for a sufficient amount of time. Once, the approximate network load is learnt, a decision can be made on the payment for the traffic.

The users declare the maximum payment per byte it is ready to make from time to time. Thus, the problem of incomplete load information at the UE can be solved. This approach reduces the cost of UEs and also avoids transmission of large volumes of control information.

*B.2. Bandwidth is a Limited Resource and Traffic has Delay Threshold*

Bandwidth is a limited resource. Section I.B.1 provides a way of abstracting the load information using LA. However, we have not yet addressed the real problem posed by Scenario B mentioned in Section I.A. Using LA, the users can decide on the payments to be made as per their requirements. However, what if there are not enough resources to support all the contesting users? This is where the competitive distribution of resources using the Auction mechanism comes in. In this approach, the UEs serve as the LA equipped bidders and the OLT serves as the auctioneer. However, the UEs are connected to the OLT via the ONUs. Therefore, ONUs collect the payments made by all the users connected to it and forwards a composite bid to the OLT. Thus, to the OLT, it appears that the ONUs are the bidders of the bandwidth.

For smooth operation, the OLT must ensure truthful bidding in order to carry out a fair allocation. Therefore, second-price auction or Vickrey auction is the obvious choice in the given situation [16]. Further, as multiple bandwidth units are to be allocated to multiple ONUs; therefore, Vickrey-Clarkes-Groves (VCG) mechanism is more suited for such an allocation procedure [17].

*C. Contributions*

To summarize, in this paper, we have proposed an Auction based opportunistic resource allocation (AuORA) protocol that performs end-to-end SLA based dynamic bandwidth allocation (DBA) of the users. The users opportunistically use the spectrum available to the PALs.

AuORA runs in conjunction with a Learning Automata based SLA aware algorithm (LASA) to optimize the way the UEs perceive the network. Hence, the UEs aid the ONUs in the bid evaluation process such that the end-to-end QoS of traffic classes involved can be maintained.

The rest of the paper is organized as follows. Section II throws light on the available literature. Section III describes the system model and the primary objectives. Section IV provides details of the proposed algorithms. Section V discusses the simulation model. Section VI showcases the results and discussion followed by the conclusion.

## II. RELATED WORK

In this section, we provide a brief overview of the available literature on SAS and also on the QoS and SLA based DBA for EPON. Thereafter, we provide a gap analysis of the available literature.

### A. Spectrum Access System

Several research works are available in the literature that is targeted towards dynamic spectrum sharing between higher priority primary users and lower priority secondary users. The authors of [18] have proposed a learning aided listen-before-talk scheme for the GAA users to access the spectrum. However, the central idea of the scheme being distributed in nature may impart interference to the PAL users if the GAA users are not able to perfectly predict PAL activity. In [19], we find a resource allocation scheme between fixed and mobile GAAs while limiting the interference to the PALs. The authors of [20] present a distributed power allocation algorithm for the GAA users so that the interference caused by the GAA users to the PALs can be minimized while at the same time the need for sharing the location information of the GAA users can be avoided. Since, the MNOs might be reluctant to share detailed location information of the BSs/users, the authors of [4] discusses a method where the mathematical distribution the lower-tier network users and the number of lower-tier transmitters are shared with the higher-tier network users. Thereby, the higher-tier network users can estimate the aggregated interference from the lower-tier network users and thereafter design exclusion zones for the lower-tier network users. In [8], we find price-based service agreements based on incumbent and licensee spectrum usage.

### B. SLA Based DBA for EPON

The authors of [21] have introduced the concept of *"excess distribution"* and they have also used the idea of *"report before data"*. A similar excess distribution-based approach that schedules after taking SLA into consideration is provided in [22]. The authors of [23] propose two algorithms to maintain QoS; Modified delay aware window sizing (M-DAWS) for high-priority traffic and delay aware grant sizing (DAGS) for medium-priority traffic. They further propose an SLA aware differential polling (DP) algorithm. DP algorithm divides the user into multiple groups based on the delay bound of the highest priority packets. In another approach [24], the polling cycle is divided into two sub-cycles; the best effort polling cycle and the real time polling cycle. A P2P live-streaming application aware architecture and protocol is proposed in [25]. The authors of [26] propose a fair excess-dynamic bandwidth allocation (FEX-DBA) that is based on a network utility maximization model. FEX-DBA is an online scheduling algorithm that works efficiently in a stand-alone EPON. Interestingly, an attempt has been made to incorporate auction in the DBA mechanism of EPON as we can find in [27]. In [27], a first price auction has been designed to promote a fair bandwidth sharing among the participating ONUs. The authors of [28] propose an evolutionary game-based approach to share the PON backhaul between two competing BSs. They argue that fixed allocation of PON resources is inefficient while satisfying the dynamic requirements of the BSs. However, they do not consider the traffic of the individual UEs.

In [29], a dynamic bandwidth allocation algorithm with demand forecasting has been proposed. The proposal reduces the end-to-end delay by using statistical modelling to forecast future bandwidth demands in a 10-gigabit-capable passive optical network. A resource management procedure for optimizing the allocation of GPON resources based on the dynamic adjustment of the SLA parameters according to estimated customer traffic patterns has been proposed in [30]. The work utilizes clustering analysis to segregate users according to their network uses based on real-time and historical data. A joint bandwidth and queue management mechanism for upstream SLA-Oriented QoS in Multi-Tenant and Multi-Service PONs is explored in [31]. The authors of [32], propose a double fair dynamic bandwidth allocation scheme based on user satisfaction. The proposal accommodates fairness both in terms of wavelength allocation and time-slot allocation.

*C. Gaps in Literature*

The available literature on SAS or EPON considers just the wireless or optical network part respectively. The works do not take into account the circumstances that may arise if the wired backhaul fails to forward the data received by the wireless BSs. Moreover, the works on SAS consider distributed access mechanisms for the GAA users. It is well-known that distributed systems must adhere to strict co-ordination in order to function efficiently. In a WOIN, a centralized spectrum allocation methodology is often useful for providing end-to-end service guarantee. Similarly, in the EPON literature, most of the papers mainly focus on either on QoS enhancement or at providing SLA based treatment. However, to the best of our knowledge, end-to-end opportunistic scheduling in WOIN/EPON has not been considered in the literature. Therefore, we come up with a new end-to-end scheduling strategy in WOIN that can satisfy SLA requirements of SAS.

## III. SYSTEM MODEL AND OBJECTIVES

In this section, we illustrate the system model and declare the functionalities of different components of the network (Fig. 3). The OLT performs bandwidth allocation in the EPON and the eNodeB is responsible for bandwidth distribution in the LTE network. Thus, the OLT with the help of the eNodeB performs the functions of the SAS controller. The SAS Controller is supported by the intelligent users that employs LA. The primary objectives of the proposal are described in the following sub-sections.

### A. End-to-end QoS Maintenance

Our proposal enforces SLA in order to provide end-to-end QoS. During connection establishment, the ONU and the users agree upon an SLA in exchange of a base fare. SLA is defined in terms of average bit rate. If a user operates within the decided SLA, the ONU must provide enough bandwidth to the user to meet the SLA. Finally, if the ONU fails to provide transmission opportunity for the SLA packets of a certain user, the ONU has to pay a penalty to the user. In this work, we assume that the PAL users have an SLA with the possibility of trying to transmit more data than their SLA. On the other hand, the GAA users have no SLA guarantees. They transmit traffic opportunistically.

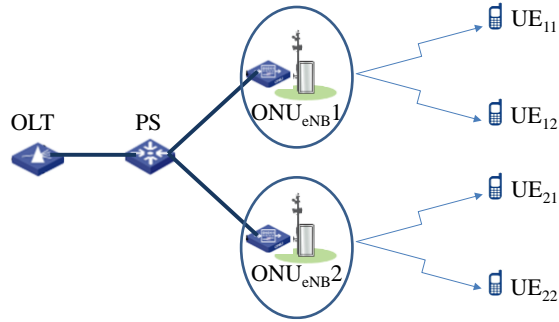

Fig. 3 System Model

### B. Opportunistic Bandwidth Allocation

Opportunistic bandwidth allocation takes place in the EPON side of the network. After receiving the packets from the users, the ONUs are responsible for forwarding the packets to the OLT. Unfortunately, the channel between the ONU and the OLT also has finite bandwidth and it may so happen that the aggregated bandwidth requirement from the ONUs is more than the available bandwidth. In such a situation, the ONU ensures the transmission of the packets (bytes) within SLA (SLA packets (bytes)). Thereafter, ONU tries to forward the packets that are above the SLA (ASLA packets (bytes)). When the bandwidth demand for the ASLA packets is higher than the available bandwidth, the bandwidth is provided on a competitive basis with the help of an auction. The auction is held only on the bandwidth available after the transmission of the SLA packets. Bandwidth is allocated in chunks, i.e., the total bandwidth is divided into finite number of chunks. Therefore, we have a multiple item (chunk) auction with multiple bidders (ONUs). The OLT allocates the chunks to the highest bidders.

In this paper, the packets from the PAL users are classified as SLA or ASLA packets depending on the volume of their traffic transmission over a finite observation window. However, the packets from the GAA users are always marked as ASLA packets as they don't have any SLA.

## IV. THE END-TO-END SCHEDULING

In this section, we describe our proposed packet classification and assignment algorithm (PCA), LASA and AuORA method in details. LASA is a LA based algorithm whereas AuORA is an auction based opportunistic bandwidth allocation mechanism. Both LASA and AuORA can be used for single as well as mixed traffic. Further, if not mentioned otherwise, we shall denote both the PAL and the GAA UEs as UEs from this point onwards.

### A. Packet Classification and Assignment algorithm

#### A.1. Packet Classification at the eNodeBs.

In the LTE side, the eNodeBs employ PCA. A separate instance of PCA is executed by the eNodeB for every single UE. PCA directs that the UE $i$ will pay a base price per byte ($\chi_i$) till its uplink transmission is within the SLA. However, as the bandwidth requirements of the UE exceed its SLA, the UE has to pay a higher amount per byte. This "increased rate" is a multiple of the base rate.

The $i^{th}$ UE sends the number of bytes in its buffer ($\beta_i^t$). The UE also informs the eNodeB, the maximum price it is willing to pay per byte.

The eNodeB keeps track of the number of SLA bytes that UE $i$ has transmitted over the last $\mathcal{M}$ time steps ($\varphi_i^t$). The count is only kept for the transmissions that are within the SLA, which is defined in terms of average bit rate. Let, $\Omega_i$ represent the number of bytes that should be trans-mitted by UE $i$ over the last $\mathcal{M}$ time steps to meet SLA.

$$\varphi_i^t = \sum_{j=t-\mathcal{M}}^{t-1} T_{SLA,i}^k \, ; \, \Omega_i = (\mathcal{M} + 1)\Gamma_{SLA,i} \tag{1}$$

TABLE I. SYMBOLS USED FOR PACKET CLASSIFICATION AND ASSIGNMENT

| SYMBOLS | DESCRIPTION |
|---|---|
| $\Gamma_{SLA,i}$ | The number of bytes that should be transmitted by UE $i$ over a single time slot to meet SLA. |
| $\beta_{SLA,i}^t$ | Number of bytes in UE $i$'s buffer in the $t^{th}$ time step that are marked as within SLA. |
| $\beta_{ASLA,i}^t$ | Number of bytes in UE $i$'s buffer in the $t^{th}$ time step that are marked as above SLA. |
| $\beta_i^t$ | Number of bytes in UE $i$'s buffer in the $t^{th}$ time step $(\beta_i^t = \beta_{SLA,i}^t + \beta_{ASLA,i}^t)$. |
| $P_i$ | Payment per byte to the eNodeB for the ASLA packets by UE $i$ (learned price). |
| $\bar{P}$ | $\bar{P} = \max_i P_i + 1$. |
| $T_{SLA,i}^t$ | Number of bytes within SLA that has been transmitted by UE $i$ in the received window in the $t^{th}$ time step. |
| $T_{ASLA,i}^t$ | Number of bytes above SLA that has been transmitted by UE $i$ in the received window in the $t^{th}$ time step. |
| $T_i^t$ | Number of bytes that has been transmitted by UE $i$ in the received window in the $t^{th}$ time step $(T_i^t = T_{SLA,i}^t + T_{ASLA,i}^t)$. |
| $\varphi_i^t$ | Number of bytes that UE $i$ has transmitted over the last $N$ time steps till the $t^{th}$ time step. |
| $\Omega_i$ | Number of bytes that should be transmitted by UE $i$ over the last $\mathcal{M}$ time steps plus current time step to meet SLA. |
| $S$ | Buffer states. |
| $C_i$ | UE $i$'s buffer capacity. |
| $\chi_i$ | Base price for SLA packets for UE $i$. |
| $w_{ij}$ | The value associated with the $i^{th}$ UE and the $j^{th}$ RC. |
| $q_{ij}$ | The maximum possible number of bytes that the $i^{th}$ UE can send on the $j^{th}$ RC. |
| $\rho_{i,j}$ | The payment received from the UE $i$ for the $j^{th}$ RC. |
| $N$ | The number of UEs. |
| $M$ | The number of RCs. |
| $\psi_i$ | Number of SLA bytes that will be dropped by UE $i$ if it is not scheduled in the current TTI. |

where, $\Gamma_{SLA,i}$ is the number of bytes that should be transmitted by UE $i$ over a single time slot to meet SLA and $T_{SLA,i}^k$ is the number of SLA bytes that has been transmitted by UE $i$ in the $k^{th}$ time step. Thus, in every time step, $(\varphi_i^t)$ plus the number of bytes present in the UE $i$'s buffer $(\beta_i^t)$ are checked against $(\Omega_i)$. This step leads to the evaluation of number of SLA and ASLA bytes for UE $i$ ($\beta_{SLA,i}^t$ and $\beta_{ASLA,i}^t$ respectively). This may give rise to three cases:

1. $\varphi_i^t \geq \Omega_i$ – Here, all the bytes in the buffer are marked as ASLA packets $(\beta_{SLA,i}^t = 0; \beta_{ASLA,i}^t = \beta_i^t)$.

2. $\varphi_i^t + \beta_i^t > \Omega_i$ and $\varphi_i^t < \Omega_i$ – Here, $\left(\beta_{SLA,i}^t = \Omega_i - \varphi_i^t\right)$ bytes in the buffer are marked as within SLA and the rest of the bytes are marked as ASLA $\left(\beta_{ASLA,i}^t = \beta_i^t - \beta_{SLA,i}^t\right)$.

3. $\varphi_i^t + \beta_i^t \leq \Omega_i$ – In this case, all the bytes present in the buffer are marked as within SLA $\left(\beta_{SLA,i}^t = \beta_i^t; \beta_{ASLA,i}^t = 0\right)$.

Once, the number of ASLA bytes $\left(\beta_{ASLA,i}^t\right)$ is calculated, the next step is to calculate the buffer states. The buffer states are normalized value of the $\beta_{ASLA,i}^t$ w.r.t buffer capacity and is found by $ceil\left(\frac{\beta_{ASLA,i}^t}{C_i}|S|\right)$, where $|S|$ denotes the number of buffer states and $C_i$ is the buffer capacity. Depending on the current state, a payment value is being evaluated. The payment value $(P_i)$ is probabilistically decided by using the input from the LASA algorithm (to be discussed in Section IVB). The selected payment value acts as the input to the resource allocation unit of the eNodeB.

*A.2. Assignment based scheduling by the eNodeB.*

In this sub-section, we describe the assignment-based procedure that we employ to schedule the UEs. In this phase, the eNodeB maximizes the earnings in the process of scheduling. We adopt the primary principle of bandwidth scheduling from our previous works [33][34]. As in [33][34], our primary target is to maximize the MAC throughput while enhancing a secondary target, which in our case is the payment from the UE.

---

**Algorithm 1 PCA Algorithm**

**Input**: $\chi_i, \beta_i^t, \varphi_i^t, \Omega_i, C_i, |S|, q_{i,j}, \overline{\mathfrak{R}_s^t}$
**Output**: $P_i$
1. *Classify packets;*
2. *Identify buffer state 's';*
3. *Select action 'a' from the set $A_s$;*
4. *Execute scheduling;*
5. *if UE i is scheduled*
6.   *Buffer the incoming packets;*
7.   *Calculate the received price;*
8. *else*
9.   *Call LASA (Algorithm 2);*
10. *end if;*

*\* PCA is operated at the eNodeB module of the $ONU_{eNB}$ every TTI (1 ms)*

---

First, the entire bandwidth is broken into discrete non-overlapping bandwidth units called re-source chunks (RCs). One RC can be allocated only to a single UE. Therefore, the eNodeB first calculates the number of bytes that can be transmitted by a UE over each of the RCs provided the RC is being allocated to the UE. Logically, this value is the minimum of the number of bytes present in the UE buffer and the transmission of number of bytes that the RC physically allows after considering the channel conditions. This calculation leads to the formation of the traffic matrix $W$. The number of bytes that can be transmitted is given by (2).

$$w_{i,j} = min(q_{ij}, \beta_i) \tag{2}$$

where, $w_{i,j}$ is the element of traffic matrix $W$ that corresponds to the $i^{th}$ UE and the $j^{th}$ RC; $q_{ij}$ is the maximum possible number of bytes that the $i^{th}$ UE can send on the $j^{th}$ RC and $\beta_i$ is the number of bytes present in $i^{th}$ UE's buffer.

Thereafter, each of the elements of the $W$ matrix is multiplied with the payment per byte that is to be received from the UE if the UE is allocated that RC. Hence, three distinct cases may arise for which the payments to be received from the UE $(\rho_{i,j})$ are calculated as:

1. If $\varphi_i^t \geq \Omega_i$ then $\rho_{i,j} = P_i w_{i,j}$.
2. If $\varphi_i^t + w_{ij} > \Omega_i$ and $\varphi_i^t < \Omega_i$ then $\rho_{i,j} = (\Omega_i - \varphi_i^t)\chi_i + (w_{ij} - \Omega_i + \varphi_i^t)P_i$.
3. If $\varphi_i^t + w_{ij} \leq \Omega_i$ then $\rho_{i,j} = \chi_i w_{i,j}$.

Please note that the above cases are similar as the ones described in Section IVA.1. Here, the distinction is that the total received payments are estimated based on the number of bytes that can be transmitted, whereas in Section IVA.1, we illustrate the packet classification procedure into SLA and ASLA packets.

Moreover, we mention that the assignment problem of network allocation requires equal number of elements in both the partites, i.e., the number of UEs $(N)$ should be equal to the number of RCs $(M)$. However, in most of the cases, $N > M$. Therefore, in order to make the cardinality of both the partites same, $(N - M)$ dummy RCs are added. Any UE assigned to a dummy RC will not receive any bandwidth in the current cycle.

Finally, it is essential that the UEs that are still within their SLA and their SLA packets have a risk of being dropped should be given a preferential treatment in the scheduling. Therefore, we ensure that if a user having SLA packets is not allocated in the current cycle and some of its packets are dropped, the eNodeB has to give a penalty. In graphical terms, we make the weight of connecting the UEs to the dummy RCs as the penalty value. The penalty is negative of the number of SLA bytes that will be dropped by UE $i$ ($\psi_i$) multiplied by a number larger than the highest payment level ($\bar{P}$). We have taken $\bar{P} \in Z^+$ as $\bar{P} = \max_i P_i + 1$. This approach is taken to ensure than an SLA user is given more priority than an ASLA user. Once the entire matrix is formulated, the eNodeB runs a maximization algorithm (ILP 3-8) to evaluate the allocation.

$$Maximize \sum_i \sum_j \alpha_{ij} \rho_{ij} - \sum_i \gamma_i \bar{P} \psi_i \tag{3}$$

Subject to,

$$\sum_i \alpha_{ij} = 1, \forall j \tag{4}$$

$$\sum_j \alpha_{ij} \leq 1, \forall i \tag{5}$$

$$\gamma_i + \sum_j \alpha_{ij} = 1, \forall i \tag{6}$$

$$\alpha_{ij} = 0 \text{ or } 1 \tag{7}$$

$$\gamma_i = 0 \text{ or } 1 \tag{8}$$

where, $\rho_{i,j}$ is the payment that the UE $i$ will make to the eNodeB if UE $i$ is allocated to RC $j$; $\alpha_{i,j}$ is a binary variable, which is equal to 1 if UE $i$ is allocated to RC $j$. Otherwise, $\alpha_{i,j} = 0$. The constraint (4) implies that an RC can be allocated to only a single UE. The constraint (5) signifies that one UE can get at most one RC. Constraint (6) ensures that UE $i$ is either allocated an RC ($\sum_j \alpha_{ij} = 1$) or the UE is mapped to the dummy RC ($\gamma_i = 1$). Constraint (7) and (8) indicates that $\alpha_{ij}$ and $\gamma_i$ are binary variables.

The ILP (3-8) is further converted into an assignment problem by following the procedure shown in [34]. Thereafter, we use the well-established *Hungarian algorithm* for the purpose of solving the assignment problem optimally in polynomial time. The PCA algorithm is summarized in Algorithm 1.

B.  *LA based SLA aware algorithm at the eNodeB[4].*

In this sub-section, we describe how an eNodeB updates the probability of choosing the payment values using the LA algorithm [7][35]. As already mentioned, we allow multiple levels of payment values according to the state of the queue. We observe that including a single level of payment value may not capture the network load conditions properly. For example, if the network load is relatively low but one UE requires more bandwidth than its prescribed SLA, it is not fair to charge a huge amount from the concerned UE. On the other hand, if there are several UEs contending for the extra bandwidth, a competition among the concerned UEs is the only way to award bandwidth to the UE that is willing to pay more. Hence, we have included more than one levels of rates over the base price.

---
[4] We remind the readers that the MEC servers situated in an eNodeB execute the LA algorithm on behalf of the UEs.

TABLE II. SYMBOLS USED IN LEARNING AUTOMATA

| SYMBOLS | DESCRIPTION |
|---|---|
| $A_s$ | Set of available actions (set of price levels) in state $S$. |
| $p_{a,i}^t$ | Probability of choosing action $a \in A_s$ in the $t^{th}$ time step for UE $i$. |
| $\overline{\mathfrak{R}_{s,i}}$ | Probability vector for $A_s$ for UE $i$. |
| $R_{a,i}$ | Reward after taking action $a$ for UE $i$. |
| $L_{a,i}$ | Penalty after taking action $a$ for UE $i$. |
| $B$ | Variable that record success $(B=0)$ or failure $(B=1)$ |

We employ LASA for the UEs to decide which of the available price levels to use. As the network operation progresses, the instances of the LASA algorithm running at the eNodeB converge to the optimal pricing option. Therefore, if the overall network load is low, there will be fewer number of competing UEs and hence a lower price level will be chosen for the UE. On the other hand, in heavy network loads, a higher price level is more likely to be chosen.

The LA algorithm runs only for the $\beta_{ASLA}^t$ packets[5]. The descriptions of the variables used in the LA are given in TABLE II. The initial probability of choosing all the actions from the action space $A_s$ are equally likely. Hence, the initial condition for the LA algorithm is given by (9).

$$p_a^1 = \frac{1}{|A_s|}, \quad \forall a \in A_s \tag{9}$$

where, $p_a^t$ is the probability of choosing action $a$ in the $t^{th}$ time step, $|A_s|$ denotes the number of available actions in a given state $S$.

During the course of learning, the agent updates $p_a^t \ \forall \ a \in A_s$ at each time step $(t)$ based on the outcome of the last decision $(B = 0 \ (success) \ or \ 1(failure))$ from (13). The outcome is the reward or penalty $(R_a \ or \ L_a)$ received by the UE after choosing the action $a$. The outcome of the decision depends on the requested and the received bandwidth from the ONU and also on wheth-

---
[5] Please note that we are discussing about UE $i$ for the $t^{th}$ time interval. Therefore, if there is no ambiguity, we will be dropping the suffix $i$ in the following discussion.

er the ONU is able to forward the received packets to the OLT. As a result, two cases arise –
1) *The UE gets the requested bandwidth* – In this case, the UE gets the requested bandwidth for the ASLA bytes from the ONU. As a result, the packets are transmitted to the ONU. However, the packets received by the ONU may or may not be further forwarded to the OLT depending upon the load condition in the EPON. If the EPON load is excessive, the packets are dropped at the ONU if the packets overshoot their delay deadline. Hence, in a congested EPON, a packet transmitted by the UE may not reach the OLT. On the other hand, if the EPON load is low, the ONU will successfully forward all the packets to the OLT. Therefore, to provide end-to-end packet delivery information, the ONU provides the update for this set of packets after it successfully forwards to the OLT or drops them. In such a situation, the probability update is performed only after receiving the packet drop information from the ONU.
2) *The UE does not get the requested bandwidth* – If a UE is not allocated enough bandwidth to transmit any ASLA packets in the present TTI, the probability update is performed immediately.

If all the ASLA packets reach successfully to the OLT, the outcome is considered as a success event and reward is generated. Otherwise, the outcome is considered as a failure and penalty is generated. Let, $T^t$ be the number of bytes from the $t^{th}$ time interval that can be transmitted by the ONU and $P^t$ be the payment per byte in the $t^{th}$ time interval. The value of $R_a$ and $L_a$ are given as:

$$R_a = \frac{\beta^t_{ASLA}}{P^t}, \qquad\qquad if\ T^t \geq \beta^t_{ASLA} + \beta^t_{SLA} \qquad (10)$$

$$L_a = \begin{cases}(T^t - \beta^t_{SLA} - \beta^t_{ASLA})P^t, & if\ \beta^t_{ASLA} + \beta^t_{SLA} > T^t \geq \beta^t_{SLA} \\ -\beta^t_{ASLA}P^t, & otherwise\end{cases} \qquad (11)$$

For the purpose of the algorithm, the Reward (and penalty) value $(R_a)$ is normalized following (12).

$$R_{a,norm} = \frac{R_a - R_{a,min}}{R_{a,max} - R_{a,min}} \qquad (12)$$

where, $R_{a,min} = 0$ is the minimum possible value of $R_a$ and $R_{a,max} = \beta^t_{ASLA}$ is the maximum possible value of $R_a$. Similarly, the penalty $(L_a)$ is also normalized with $L_{a,min} = -\beta^t_{ASLA}\bar{P}$ and $L_{a,max} = 0$. We use this normalized value to update the values of $p^t_a$ in (13).

$$\text{If } B = 0:$$
$$p^{t+1}_a = p^t_a + R_{a,norm}[1 - p^t_a],$$
$$p^{t+1}_j = (1 - R_{a,norm})p^t_j, \forall j \neq a$$
$$\text{If } B = 1:$$
$$p^{t+1}_a = (1 - L_{a,norm})p^t_a,$$
$$p^{t+1}_j = \frac{L_{a,norm}}{1-|A_s|} + (1 - L_{a,norm})p^t_j, \forall j \neq a \quad (13)$$

In our case, we have multiple UEs (agents). Every single agent takes its actions independently. Moreover, any UE$(i)$ has no idea about the actions taken by UEs$(j)$ where $j \neq i$. The UE$(i)$ considers all the outcomes due to the actions taken by the other UEs as environmental factor. UE$(i)$ therefore gets an abstract view of the environment conditions without having to interact with any other UE.

---

### Algorithm 2 LASA

**Input**: $\beta^t_{ASLA,i}, \beta^t_{SLA,i}, T^t_i, \mathfrak{R}_{s,\iota}$
**Output**: $\overline{\mathfrak{R}_{s,\iota}}$
1. If $T^t_i \geq \beta^t_{ASLA,i} + \beta^t_{SLA,i}$;
2.      Calculate $R_{a,i}$;
3.      Calculate normalized value of $R_{a,i}$;
4.      $B = 0$;
5. else
6.      Calculate $L_{a,i}$;
7.      Calculate normalized value of $L_{a,i}$;
8.      $B = 1$;
9. Update action probability vector $\overline{\mathfrak{R}_{s,\iota}}$, i.e., $p^{t+1}_{a,i} \forall a \in A_s$ according to (13);

*LASA is operated at the eNodeB module of the ONU$_{eNB}$*

---

C. *Auction based opportunistic resource allocation protocol*

C.1. *An offline protocol.*

AuORA is an offline protocol as it considers the information from all the ONUs while taking the scheduling decision. It follows report before data format, i.e., report is transmitted at the be-ginning of the transmission slot [21]. This approach helps in reducing the idle time created between cycles due to DBA processing. It is expected that if report before data is used and the nearest ONU is polled first then there will be very little idle time. We believe that other options like performing DBA just after receiving the report message from the $(\Phi - 1)^{th}$ ONU is not advisable as the results will drift from the optimum values. In this description, '$\Phi$' is the number of ONUs. For even better performance, one can schedule the ONU with the largest transmission window at the end of the cycle.

*C.2. Report Preparation.*

Bid preparation is the most essential part of the AuORA protocol. The entire responsibility of preparation of bids lies with the ONU, since they are the bidders/purchasers of bandwidth from the OLT. According to AuORA, every ONU must follow the same guidelines. The basic idea of the protocol is that if an ONU needs more bandwidth, it has to pay more.

The bids are simply created by summing up the total payment received by the eNodeBs from the UEs for the ASLA packets.

$$v_n = \Sigma_{i \epsilon \Im} P_i \qquad (14)$$

where, $v_n$ is the bid value of the $n^{th}$ ONU, $P_i$ is the payment received for packet $i$ and $\Im$ is the set of ASLA packets.

Our proposal indirectly incorporates the load scenario in the LTE network while performing scheduling in EPON. Therefore, if there is any urgent packet from the LTE side, it will come with a higher payment and will be given a preferential treatment. Further, the ONU must also ensure that it can transmit the SLA packets. Thus, once the SLA packets are transmitted, we ensure an expedited treatment to any ASLA packet seeking immediate transmission. Thus, the $n^{th}$ ONU reports the number of SLA packets, the number of ASLA packets and the bid for the bandwidth purchase. The format of the report for the $n^{th}$ ONU is $(R_{SLA,n}, R_{ASLA,n}, v_n)$, where, $R_{SLA,n}$ is the number of chunks required to transmit the SLA packets, $R_{ASLA,n}$ is the number of chunks required to transmit the ASLA packets and $v_n$ is the bid value. A chunk can be as small as a single byte. In any case, the total number of chunks available in the system is restricted by the cycle length, which is 2 ms in our case.

Since, we operate VCG auction mechanism [38]; the ONU that wins the bandwidth will al-ways pay an amount which is less than its bid. Hence, the profit of the ONU is the difference between the bid that it places and the actual payment that it makes to the OLT. Further, please note that the valuation of the ONU is equal to the bid value (which is obtained from the payment from the UEs). This happens because VCG ensures truthful bidding.

*C.3. Queue Management.*

In order to maintain proper QoS, the ONU maintains three queues. The first one is for highest priority real-time traffic (voice), the second one is for the medium priority real-time traffic (video) and the third one is for best effort traffic (data). The packets in the real-time queues are arranged in the order of their arrival.

Upon receiving a packet, we assign a weight and value to it. Typically, the weight of the $i^{th}$ packet ($w_i$) is its size and the value ($\vartheta_i$) is the payment received from the UE for that packet. These stored values serve two purposes. Firstly, they help in the Report message generation. Secondly, they help in the procedure of marking a packet for transmission.

We mention here that we do not follow strict priority scheduling where packets from a lower priority queue are transmitted only when all its higher priority queues are empty. We maximize the payment received by transmitting the most valued packets. The packets that are marked to be transmitted in the current cycle may belong to any of the three queues. Further, the SLA packets are always transmitted before the ASLA packets.

*C.4. VCG auction and bandwidth allocation.*

The responsibility of bandwidth allocation lies with the OLT. First, the OLT reserves the bandwidth required for transmission of the SLA packets. This step is taken in order to ensure that an ONU is not starved because of richer and bandwidth hungry ONUs.

Having received all the bid values ($v_n$) and reserving the bandwidth for SLA packets for all ONUs, the OLT initiates the allocation procedure for the excess bandwidth.

Please note that the allocation is essentially a fractional knapsack problem. In a fractional knapsack problem, whole items are inserted in the knapsack as long as possible. Finally, when there is not enough capacity left in the knapsack to insert a whole item, a fraction of the item is inserted into the knapsack. The problem is defined as,

$$Maximize \sum_n \alpha_n v_n \qquad (15)$$

Subject to,

$$\sum_n \alpha_n R_{ASLA,n} \leq C \quad (16)$$

$$0 \leq \alpha_n \leq 1 \quad (17)$$

where, $C$ is the bandwidth left after taking care of the SLA packets, $R_{ASLA,n}$ is the requirement of the $n^{th}$ ONU and $v_n$ is the price that the $n^{th}$ ONU is prepared to pay on receiving the allocation. $\alpha_n$ is a decision variable where, $\alpha_n = 0$ indicates that the $n^{th}$ ONU is not allocated any bandwidth. On the other hand, $0 < \alpha_n \leq 1$ indicates allocation.

TABLE III. SYMBOLS USED IN REPORT PREPARATION AND VCG AUCTION

| SYMBOLS | DESCRIPTION |
|---|---|
| $\mathfrak{J}$ | Set of ASLA packets. |
| $R$ | Number of requested bytes. |
| $R_{ASLA,n}$ | Requirement of the $n^{th}$ ONU for the ASLA packets. |
| $R_{ASLA,n}$ | Requirement of the $n^{th}$ ONU for the SLA packets. |
| $v_n$ | Bid value of the $n^{th}$ ONU. |
| $k_n^*$ | Optimal allocation when the $n^{th}$ ONU is present. |
| $k_{-n}^*$ | Optimal allocation when the $n^{th}$ ONU is absent. |
| $\wp(k)$ | Total price of the Bandwidth chunks as per allocation. |
| $\tau_n$ | VCG discount for the $n^{th}$ ONU. |
| $\mathcal{H}_n$ | Payment made by the $n^{th}$ ONU. |
| $\varpi_n$ | Number of bandwidth chunks allocated to the $n^{th}$ ONU in the VCG auction for the ASLA packets. |
| $\vartheta_{ij}$ | The value associated with the $i^{th}$ packet of the $j^{th}$ type. |
| $G$ | The allocated bandwidth. |

It is well known that a fractional knapsack problem can be optimally solved using a greedy algorithm after sorting the items in a decreasing order of the ratio $\left(\frac{v_n}{R_{ASLA,n}}\right)$. Then, the allocation is done following the evaluated order.

Once, the allocation is completed, the next step is to receive the payment from an allocated ONU in exchange of the number of chunks. The payment evaluation is computed according to the principles of VCG auction mechanism. VCG is a second price auction and therefore, the concerned ONU (an agent in auctioning terms) has to pay a sum which is lower than its bid value (discounted value) [36]. The idea behind the discount is that an allocated agent makes a contribution to the system. If the agent was not present in the system, the system would have collected lesser profit. Hence, the agent is given a discount over its bid value that is equal to the profit that is earned by the system due to the presence of the agent in the system. This discount is the profit of the ONU in the packet transmission.

Let $k_n^*$ and $k_{-n}^*$ be the optimal assignments when the $n^{th}$ ONU is present and absent respectively. Similarly, let $\wp(k_n^*)$ and $\wp(k_{-n}^*)$ be the prices of the resources as per the bid values. Therefore, the profit brought by $n^{th}$ ONU is given by (18).

$$\tau_n = \wp(k_n^*) - \wp(k_{-n}^*) \tag{18}$$

Hence, the auctioneer gives a discount of $\tau_n$ to $n^{th}$ ONU and the price that the $n^{th}$ ONU needs to pay is given by (19).

$$\mathcal{H}_n = \left(\frac{v_n}{R_{ASLA,n}}\right)\varpi_n - \tau_n \tag{19}$$

Where, $\varpi_n$ is the number of chunks allocated by the VCG auction to the $n^{th}$ ONU.

*C.5. Packet transmission.*

This sub-section describes the policy adopted for marking a packet so that it can be transmitted in the current cycle. First, the SLA packets are marked for transmission and the bandwidth required for their transmission is deducted from the total available bandwidth. Thereafter, the ONU selects the packets having the highest payments from the three queues. In other words, the ONU seeks to maximize the profit earned by sending its buffered packets by fitting them into the remaining bandwidth. This requirement matches with that of a knapsack problem. The problem is formally defined as follows.

$$Maximize \sum_i \sum_j \alpha_{ij} \vartheta_{ij} \tag{20}$$

Subject to,

$$\sum_i \sum_j \alpha_{ij} w_{ij} \leq G \tag{21}$$

$$\alpha_{ij} = 0 \text{ or } 1 \tag{22}$$

where, $j \in \{voice, video, data\}$, $\alpha_{i,j}$ is a binary variable indicating the scheduling information of the $i^{th}$ packet of the $j^{th}$ type, $\vartheta_{ij}$ is the value associated with the $i^{th}$ packet of the $j^{th}$ type, $w_{i,j}$ is the size of the $i^{th}$ packet of the $j^{th}$ type and $G$ is the allocated bandwidth.

In order to minimize the execution time of the algorithm, the knapsack algorithm is executed with the help of greedy algorithm. Greedy algorithm works perfectly when there is divisible item in the knapsack (fractional knapsack). Unfortunately, in EPONs, a packet cannot be divided and hence, knapsack should ideally be solved using dynamic programming. However, using dynamic programming makes the processing of the problem beyond practical limits. Therefore, even though greedy gives slightly sub-optimal results, we use it for solving the knapsack. We believe that the ONU will have finite number of packets to deal with while solving the knapsack and hence the processing time complexity will remain within practical limits.

Algorithm 3 AuORA

**Input** : $P_i$
**Output** : $T^t$
1. *Prepare Bids*;
2. *Receive Grant from OLT;*
3. *Check for delay deadlines of the packets*;
4. *if packet delay is within deadline*
5.    *Forward packets to OLT if Bandwidth is available*;
6.    *Update $T^t$*;
7. *else*
8.    *Drop packet*;
9. *end if*;
10. *if all packets received from a particular UE in a single TTI are either forwarded of dropped*;
11.    *Call LASA (Algorithm 2)*;
12. *end if*;

*\* AuORA is operated at the ONU module of the ONU$_{eNB}$ every EPON cycle (2 ms)*

*C.6. Reward/Penalty Feedback.*

This sub-section deals with the details of how the ONU gives a rewards/penalty feedback to the LASA module of the ONU in order to aid the LA algorithm. As we have already mentioned, a packet received over the wireless interface may get dropped in the EPON. In such a situation, the ONU must return the payment received from the concerned UE and provide a feedback such that the UE becomes more successful in the subsequent attempts. Therefore, ONU records the information about the status of the packets after they are transmitted or dropped. The ONU feeds this value to the LA framework with the correct action (bid value) so that the action probabilities ($p_a^{t+1}$) can be updated for the next time slot. The steps of AuORA are listed in Algorithm 3.

V.   SIMULATION MODEL

The simulation model has been developed using OMNeT++ network simulator. The EPON has been built according to the guidelines given in [9]. The parameters used in the simulation are summarized in TABLE IV.

TABLE IV. EPON PARAMETERS [9]

| PARAMETER | VALUES |
|---|---|
| Propagation delay | uniformly distributed between [50,100] µs |
| Link speed | 1 Gbps |
| No. of ONUs | 16 |
| Guard Interval | 5 µs |
| Max. cycle time for Limited IPACT | 2 ms |
| ONU buffer capacity | 10 mega bytes |

TABLE V. LTE PARAMETERS [37][39]

| PARAMETER | VALUE |
|---|---|
| Scenario | Uma |
| Inter-site Distance | 500 m |
| System Bandwidth | 10 MHz |
| Center Frequency | 2 GHz |
| No. of PRBs ($n_{PRB}$) | 50 (48 for data) |
| No. of PRBs in a RC | 6 |
| Path loss ($PL$) Model | Non Line of Sight |
| Shadowing Standard Deviation | 4 dB |
| UE Max Transmit Power ($P_{max}$) | 24 dBm |
| Uplink Power Control (PC) | $max\ (P_{max}, P_o + \alpha P + 10 \log_{10} n_{PRB})$ |
| $\alpha$ for PC | 1.0 |
| $P_o$ for PC | -106 dBm |
| UE Distribution | Poisson Point Process (PPP) |
| Traffic Model | Two State Markov VoIP |
| Simulation duration | 10 seconds (10,000 iterations) |
| MCS | QPSK, 16-QAM, 64-QAM with varying code-rates as given in [37] |

A seven-cell scenario has been considered in the LTE network, where a single cell is surrounded by six first tier cells. The work assumes that the center cell is the serving cell and the first-tier cells provide the interfering signal power. All the measurements have been taken in the center cell and only the packets coming from the center cell is fed to the EPON. However, as the throughput from the center cell is limited to 10 Mbps, we have used 90% of the EPON link for background traffic and only 10% of the EPON capacity is used for backhauling the LTE traffic. The UEs have been deployed in the cells by following a Poisson point process. Uplink transmission power control and MCS have been considered as per the guidelines provided in [39]. Block fading channel model has been used, where the channel conditions remain constant over a TTI [40]. The details of the simulation parameters are listed in TABLE V.

We have used a single class of traffic for illustrating the performance of AuORA. However, the simulation can be easily extended to multiple traffic with varying priorities.

### A. Benchmarks

For the comparisons, *Dynamic Hungarian Algorithm with Modification (DHAM)* [33] has been used in the LTE network. We have used the combination of DHAM because to the best of our knowledge, it is the most efficient uplink scheduling protocol available in LTE.

For the EPON Network we have used two different protocols for the comparative studies –

- *Interleaved Polling with Adaptive Cycle Time* [9] – IPACT is one of the most famous and standard protocol used in the EPON uplink. Therefore, it is one of our automatic choices for the benchmarking.

- *Fair excess-dynamic bandwidth allocation* [26] – To the best of our knowledge, FEX-DBA is the most recent protocol available in the literature that deals with SLA based fair scheduling protocol for EPON uplink.

## VI. RESULTS AND DISCUSSIONS

### A. End-to-end Performance

The performance of the AuORA algorithm working in tandem with LASA has been evaluated and discussed in this sub-section. Through simulations, we find that the LASA along with the AuORA (LASA-AuORA) algorithm transmits majority of the SLA packets even when the network load is high. The result is clearly demonstrated in Fig. 4a and Fig. 5a. This happens because AuORA ensures that the SLA packets are first transmitted in every cycle. However, we observe a very low percentage of SLA packets being dropped at heavy network loads because the packets have already suffered high delay in the LTE network and cannot wait for even the very first transmission opportunity in the EPON. On the other hand, IPACT and FEX-DBA do not explicitly reserve bandwidth for the SLA packets and hence, show lesser efficiency when it comes to delivery of SLA packets. The same trend is observed when SLA packet arrival rate is 10% and 50% of the total network capacity.

Whereas, the SLA packets are accommodated at the expense of the ASLA packets and as a result, LASA-AuORA drops higher number of ASLA packets than IPACT and FEX-DBA (Fig. 4b and Fig. 5b). Here, we should note that the ASLA packets correspond to the on-demand applica-tions. However, we emphasize that our priority is to successfully transmit the SLA packets. Thereafter, the ASLA packets outside SLA are served on a competitive basis. As a result, all the users might not receive enough transmission opportunity for their ASLA packets and hence, the higher drop.

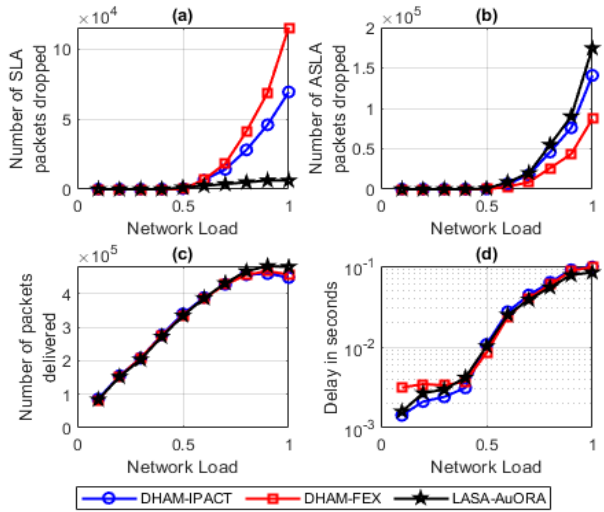 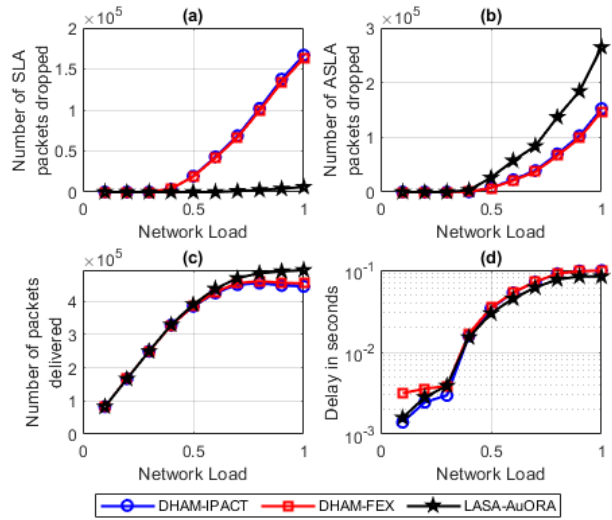

Fig. 4 End-to-end performance of the LASA-AuORA algorithm – loads of all the ONUs are varying; SLA – 10% of Network Load (a) Number of SLA packets dropped vs Network Load. (b) Number of packets forwarded vs Network Load. (c) Delay vs Network Load.

Fig. 5 End-to-end performance of the LASA-AuORA algorithm – loads of all the ONUs are varying; SLA – 50% of Network Load (a) Number of SLA packets dropped vs Network Load. (b) Number of packets forwarded vs Network Load. (c) Delay vs Network Load.

Interestingly, owing to the end-to end quality control provided by the LASA-AuORA algorithm, the overall packet delivery is more efficient. We can see this clearly from Fig. 4c and Fig. 5c. Whereas, IPACT and FEX-DBA do not take any extra measure to expedite the transmission of the packets that are on the verge of crossing the delay deadlines. As a result, they induce higher packet loss as compared to LASA-AuORA algorithm.

Similarly, as LASA-AuORA tries to minimize the number of packets that cross delay deadlines, the overall delay performance of LASA-AuORA is also better than the benchmarks (Fig. 4d and Fig. 5d).

### B. *Effect of Network Load on Learning of the UEs*

In this sub-section, we illustrate the efficiency of the LASA algorithm. First, we mark two UEs and vary their traffic load. We enable these two marked UEs with the capability of learning so as to efficiently decide on the payment per byte for their ASLA transmissions. All the unmarked UEs do not possess any learning capabilities. Hence, the unmarked UEs only pay the base price for the ASLA packets. Thereafter, we generate two scenarios –

- In the first scenario, the unmarked UEs transmit packets at 20% of the network load (low load condition).
- In the second scenario, the unmarked UEs transmit at 70% of the network load (high load condition).

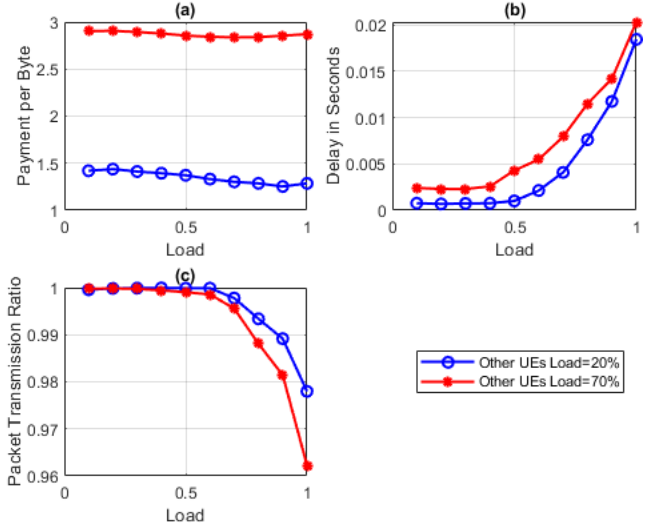

Fig. 6 Learning performance of the LASA algorithm in different network load conditions (a) Payment per Byte vs Network Load. (b) Delay vs Network Load. (c) Packet Transmission Ratio vs Network Load.

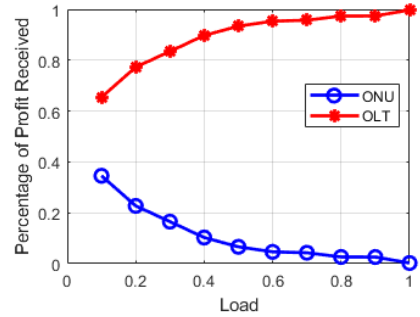

Fig. 7 Profit Sharing between OLT and ONUs.

In Fig. 6a, we observe that the learning performance is indeed dependent on the network load. Since, the overall network load is much higher in Scenario 2; the marked UEs have to pay a much higher price per bit in order to transmit their ASLA packets. The higher payment ensures comparable QoS in Scenario 1 and Scenario 2, in terms of packet delay and packet transmission ratio as can be seen from Fig. 6b and Fig. 6c. However, the delay is slightly higher in Scenario 2 because the overall network load is higher. The packet transmission ratio is slightly lower in Scenario 2 for the same reason. Please note that both the marked UEs exhibit similar average performance in terms of payment, packet delay and packet transmission ratio. Hence, we can conclude from Fig. 6 that both QoS and payment-based fairness is maintained among the users. Further, we also observe that the learning-based payment is adjusted by the competing users depending upon the overall resource availability.

## C. Profit sharing between the ONU and the OLT

In this sub-section, we illustrate the sharing of the collected credits from the UEs between the OLT and the ONUs. The ONUs simply collect the money received from the UEs and prepare the bid for their operation. The OLT allocates bandwidth in exchange of the payment. Since, an auction procedure is being used for bandwidth allocation, it is expected that the OLT will always end up with the major share of the money. The ONUs, on the other hand, receives slight profit as the second price VCG auction is used. The profit distribution can be clearly seen in Fig. 7. Note that the profit share of all the ONUs are statistically similar.

When the network load is low, the competition among the ONUs is less severe. The bandwidth demand is often lower than the availability of bandwidth slots. Since, VCG is employed, when the bandwidth demand is lower than available bandwidth, the ONUs receive the extra bandwidth free of cost. Bandwidth is allocated free of cost because the discount that is given to a player is equal to the benefit it provides to the system. In this case, if the concerned ONU is absent, the bandwidth is not utilized. Therefore, the ONU receives 100% discount. As a result, the profit share of the ONU is much higher in case of lower network loads. As the network load gradually increases, the network demand exceeds the network availability and the profit of the OLT increases.

## Conclusion

In this paper, we have proposed an end-to-end mechanism for QoS enhancement in a Wireless Optical Integrated Network that designs a medium access control layer managed intelligent spectrum access system. The scheduling in the LTE network depends on the SLA of the UEs and also on the learning influenced payments that the UE is ready to make for the packets that are above SLA. The payment received in the wireless transmission is used to prepare the bids in the VCG auction that takes place in the EPON. Thus, the UEs, through their payments have an indirect control over the scheduling in the EPON. The AuORA protocol along with the LASA protocol provides SLA guarantees. The model of the AuORA protocol is in line with the supply-demand idea of Economics. The entity that wishes more bandwidth needs to pay a higher amount to acquire it when the bandwidth demand is more than the bandwidth availability. Thus, the proposal ensures a fairness in user payment by taking the network load into consideration.

## References

[1] M.D. Mueck, S. Srikanteswara, and B. Badic. "Spectrum sharing: Licensed shared access (LSA) and spectrum access system (SAS)." *Intel White Paper* (2015): 1-26.
[2] M. M. Sohul, M. Yao, T. Yang and J. H. Reed, "Spectrum access system for the citizen broadband radio service," *in IEEE Communications Magazine,* vol. 53, no. 7, pp. 18-25, July 2015.

[3] Y. Ye, D. Wu, Z. Shu and Y. Qian, "Overview of LTE Spectrum Sharing Technologies," *in IEEE Access*, vol. 4, pp. 8105-8115, 2016.

[4] S. Srikanteswara et al. "SAS PAL GAA co-channel interference mitigation." *U.S. Patent No. 10,506,444.* 10 Dec. 2019.

[5] H. Yang, A. Alphones, Z. Xiong, D. Niyato, J. Zhao and K. Wu, "Artificial-Intelligence-Enabled Intelligent 6G Networks," *in IEEE Network,* vol. 34, no. 6, pp. 272-280, Nov.-Dec. 2020.

[6] M. Elsayed and M. Erol-Kantarci, "AI-Enabled Future Wireless Networks: Challenges, Opportunities, and Open Issues," in *IEEE Vehicular Technology Magazine*, vol. 14, no. 3, pp. 70-77, Sept. 2019.

[7] K. S. Narendra and M. A. L. Thathachar, "Learning automata: A survey," *IEEE Trans. Syst., Man, Cybern.*, vol. SMC-14, pp. 323–334, 1974.

[8] M. Mueck et al., "Methods and devices for shared spectrum allocation." *U.S. Patent No. 10,812,986.* 20 Oct. 2020.

[9] G. Kramer, B. Mukherjee, G. Pesavento, "IPACT: A Dynamic Protocol for Ethernet PON (EPON)," *IEEE Communications Magazine*, vol. 40, no. 2, pp. 74-80, Feb. 2002.

[10] G. Shen, R. S. Tucker, C-J Chae, "Fixed Mobile Convergence Architectures for Broadband Access: Integration of EPON and WiMAX," *IEEE Communications Magazine*, vol. 45, no. 8, pp. 44-50, Aug. 2007.

[11] C. Ranaweera, E. Wong, C. Lim, A. Nirmalathas, "Next Generation Optical-Wireless Converged Network Architectures," *IEEE Network*, vol. 26, no. 2, pp. 22-27, Mar.-Apr. 2012.

[12] M. A. Ali, G. Ellinas, H. Erkan, A. Hadjiantonis, R. Dorsinville, "On the Vision of Complete Fixed-Mobile Convergence," *Journal of Lightwave Technology*, vol. 28, no. 16, pp. 2243-2357, Aug. 2010.

[13] K. Ramantas, K. Vlachos, A.N. Bikos, G. Ellinas, A. Hadjiantonis, "New unified PON-RAN access architecture for 4G LTE networks," *IEEE/OSA Journal of Optical Communications and Networking*, vol. 6, no. 10, pp. 890-900, Oct. 2014.

[14] A. Mukhopadhyay, G. Das, "A Ring-Based Wireless Optical Network to Reduce the Handover Latency," *IEEE/OSA Journal of Lightwave Technology,* vol. 33, no. 17, pp. 3687-3697, Sep. 2015.

[15] N. Abbas, Y. Zhang, A. Taherkordi and T. Skeie, "Mobile Edge Computing: A Survey," in *IEEE Internet of Things Journal*, vol. 5, no. 1, pp. 450-465, Feb. 2018.

[16] W. Vickrey. "Counter speculation, auctions, and competitive sealed tenders." *Journal of Finance*, vol. 16, no. 1, pp. 8-37, Mar. 1961.

[17] E. Clarke "Multi-part pricing of public goods," *Public Choice*, vol. 11, no. 1, pp. 17–23, 1971.

[18] C. Tarver et al., "Enabling a "Use-or-Share" Framework for PAL–GAA Sharing in CBRS Networks via Reinforcement Learning," *in IEEE Transactions on Cognitive Communications and Networking*, vol. 5, no. 3, pp. 716-729, Sept. 2019.

[19] S. Basnet, Y. He, E. Dutkiewicz and B. A. Jayawickrama, "Resource Allocation in Moving and Fixed General Authorized Access Users in Spectrum Access System," in *IEEE Access*, vol. 7, pp. 107863-107873, 2019.

[20] Y. He, B. A. Jayawickrama and E. Dutkiewicz, "Distributed Power Allocation Algorithm for General Authorised Access in Spectrum Access System," *2019 IEEE Wireless Communications and Networking Conference (WCNC)*, 2019, pp. 1-6.

[21] C. M. Assi, Y. Ye, S. Dixit and M. A. Ali, "Dynamic bandwidth allocation for quality-of-service over Ethernet PONs," *IEEE Journal on Selected Areas in Communications,* vol. 21, no. 9, pp. 1467-1477, Nov. 2003.

[22] S. I. Choi and J. Park, "SLA-Aware Dynamic Bandwidth Allocation for QoS in EPONs," *IEEE/OSA Journal of Optical Communications and Networking*, vol. 2, no. 9, pp. 773-781, Sept. 2010.

[23] A. Dixit, B. Lannoo, G. Das, D. Colle, M. Pickavet and P. Demeester, "Dynamic bandwidth allocation with SLA awareness for QoS in ethernet passive optical networks," *IEEE/OSA Journal of Optical Communications and Networking*, vol. 5, no. 3, pp. 240-253, Mar. 2013.

[24] H-T. Lin, C-L. Lai and C-L. Liu, "Design and analysis of a frame-oriented dynamic bandwidth allocation scheme for triple-play services over EPONs," *Elsevier Computer Networks,* vol. 64, pp. 339-352, May 2014.

[25] A. T. Liem, I. S. Hwang, A. Nikoukar, C. Z. Yang, M. S. Ab-Rahman and C. H. Lu, "P2P Live-Streaming Application-Aware Architecture for QoS Enhancement in the EPON," *IEEE Systems Journal,* vol. PP, no. 99, pp. 1-11, 2016.

[26] N. Merayo, P. Pavon-Marino, J. Aguado, R. Durán, F. Burrull, and V. Bueno-Delgado, "Fair Bandwidth Allocation Algorithm for PONs Based on Network Utility Maximization," *Journal of Optical. Communications and Networking.* vol. 9, pp 75-86, 2017.

[27] A. R. Hedayati, M. N. Fesharaki, K. Badie and V. Aghazarian, "Arishtat: auction-based dynamic bandwidth allocation method in Ethernet passive optical networks," *IET Communications,* vol. 5, no. 15, pp. 2116-2124, Oct. 2011.

[28] I. Loumiotis, P. Kosmides, E. Adamopoulou, K. Demestichas and M. Theologou, "Dynamic Allocation of Backhaul Resources in Converged Wireless-Optical Networks," *IEEE Journal on Selected Areas in Communications,* vol. 35, no. 2, pp. 280-287, Feb. 2017.

[29] K. A. Memon, K. H, Mohammadani, A. A. Laghari, R. Yadav, B. Das, W. U Tareen, X. Xin, "Dynamic bandwidth allocation algorithm with demand forecasting mechanism for bandwidth allocations in 10-gigabit-capable passive optical network," Optik. vol. 183, pp 1032-1042, Apr. 2019.

[30] N. E. Frigui, T. Lemlouma, S. Gosselin, B. Radier, R. Le Meur and J. Bonnin, "Dynamic reallocation of SLA parameters in passive optical network based on clustering analysis," *2018 21st Conference on Innovation in Clouds, Internet and Networks and Workshops (ICIN)*, Paris, 2018, pp. 1-8.

[31] F. Slyne and M. Ruffini, "Joint Dynamic Bandwidth and Queue Management for Upstream SLA-Oriented QoS in Multi-Tenant, Multi-Service PONs," IEEE ONDM, May 2020.

[32] N. Zhan, C. Gan, J. Hui and Y. Guo, "Fair Resource Allocation Based on User Satisfaction in Multi-OLT Virtual Passive Optical Network," in *IEEE Access*, vol. 8, pp. 134707-134715, 2020.

[33] A. Mukhopadhyay, G. Das, V. S. K. Reddy, "A Fair Uplink Scheduling Algorithm to Achieve Higher MAC Layer Throughput in LTE", *in proc. of IEEE International Conference on Communications (ICC),* Jun. 2015.

[34] A. Mukhopadhyay and G. Das, "Low Complexity Fair Scheduling in LTE/LTE-A Uplink Involving Multiple Traffic Classes," in *IEEE Systems Journal*, vol. 15, no. 2, pp. 1616-1627.

[35] K. S. Narendra and M. A. L. Thathachar, Learning Automata: An Introduction. Englewood Cliffs, NJ, USA: Prentice-Hall, 1989

[36] Y. Narahari*,* "Vickrey-Clarke-Groves (VCG) Mechanisms" *in Game Theory and Mechanism Design,* IISc Lecture Notes Series: vol. 4, May 2014.

[37] Report ITU-R M.2135-1 Guidelines for evaluation of radio interface technologies for IMT-advanced.

[38] T. Groves, "Incentives in Teams". *Econometrica,* vol. 41, no. 4, pp. 617–631, Jul. 1973.

[39] 3GPP TS 36.213, "LTE: Evolved Universal Terrestrial Radio Access (E-UTRA); Physical layer procedures" (version 8.8.0 Release 8, 2009-10).

[40] M. Mehta, S. Khakurel and A. Karandikar," Buffer Based Channel Dependent Uplink Scheduling in Relay-Assisted LTE Networks", *in proc. of* IEEE WCNC'12, pp.1777-1781, Shanghai, Apr. 2012.